\documentclass[seceq]{ptptex}

\def\rhovec{\mbox{\boldmath $\rho$}}

\newcommand{\beq}{\begin{eqnarray}}
\newcommand{\eeq}{\end{eqnarray}}

\usepackage{graphicx}

\begin{document}

\title{Structure of $^{10}_{\Lambda}$Be and $^{10}_{\Lambda}$B
 hypernuclei
studied with the four-body cluster model}


\author{Emiko \textsc{Hiyama}%
, and Yasuo \textsc{Yamamoto}}

\inst{
Nishina Center for
Accelerator-Based Science,
Institute for Physical and Chemical
Research (RIKEN), Wako, Saitama,
351-0198,Japan
}

%

\abst{
The structure of the isodoublet hypernuclei,
$^{10}_{\Lambda}$B and $^{10}_{\Lambda}$Be within
the framework of an $\alpha +\alpha +\Lambda +N$
four-body cluster model is studied.
Interactions between the constituent subunits are
determined so as to reproduce reasonably well the
observed low-energy properties of the
$\alpha \alpha$, $\alpha N$, $\alpha \Lambda$,
$\alpha \alpha \Lambda$ and $\alpha \alpha N$ subsystems.
Furthermore, the two-body $\Lambda N$ interaction is
adjusted so as to reproduce the $0^+$-$1^+$ splitting
of $^4_{\Lambda}$H. 
The $\Lambda$ binding energies of  $^{10}_{\Lambda}$B 
and $^{10}_{\Lambda}$Be are 8.76 MeV and 8.94 MeV, respectively.
The energy splitting of the $1^-$-$2^-$ levels in 
$^{10}_{\Lambda}$B is 0.08 MeV, which does not contradict 
the experimental report in BNL-E930.
An even-state $\Lambda N$ charge symmetry breaking (CSB) 
interaction determined from the A=4 systems works
repulsively by $+0.1$ MeV (attractively by $-0.1$ MeV)
in $^{10}_{\Lambda}$Be ($^{10}_{\Lambda}$B).
We discuss a possibility  that an odd-state CSB
interaction improves the fitting to the experimental data
of $A=10$ double $\Lambda$ hypernuclei.}

\maketitle

\section{Introduction}

One of the primary goals in hypernuclear physics is
to extract information about baryon-baryon interactions in
a unified way.
By making use of the hyperon($Y$)-nucleon($N$) scattering data and
the rich complementary $NN$ data, several types of
$YN/YY$ interaction models have been proposed
that are  based on the SU(3) and SU(6) symmetries.
However, these $YN/YY$ interaction models have a great deal of ambiguity
at present, since the $YN$ scattering experiments are extremely limited
and there is no $YY$ scattering data.
Therefore, it is important to extract useful information on
$YN/YY$ interactions from studies of hypernuclear structure.
In the case of the $\Lambda N$ sector, the results of
 high-resolution $\gamma$-ray experiments
have been quite important for such a purpose,
where level structures of $\Lambda$ hypernuclei
are determined  within  keV systematically.

Theoretically, a powerful calculation method,
the Gaussian Expansion Method (GEM) \cite{Hiyama03}, was proposed as a means to
perform accurate  calculations of the structure
for three- and four-body system.
GEM has been used to successfully study structures for a variety of few-body
systems in atomic, baryonic and quark-level problems.
In order to extract  information about the  $\Lambda N$ interaction,
this method was applied to $s$- and $p$-shell $\Lambda$-hypernuclei
represented by three- and/or four-body models composed of $\Lambda$
and nuclear-cluster subunits, and the spin-dependent parts of
the $\Lambda N$ interactions were determined using
the results of the $\gamma$-ray experiments:
In Ref.\citen{Hiyama00}, the $\Lambda N$ spin-orbit interactions were determined
from the observed energies of spin-doublet states in
$^{9}_\Lambda$Be ($^{13}_\Lambda$C) represented by the
$\alpha \alpha \Lambda$ ($\alpha \alpha \alpha \Lambda$) cluster model.
In Ref.\citen{Hiyama06}, the $\Lambda N$ spin-spin interactions in even- and odd-states
were investigated through the combined analyses for
$^4_\Lambda$H ($^4_\Lambda$He) and $^7_\Lambda$Li ($\alpha p n \Lambda$),
where the above spin-orbit interaction was used as an input.
These works indicate that we are now entering a new stage to extract
detailed information on the $\Lambda N$ interaction by combining few-body
calculations and $\gamma$-ray experimental data.

In this work, on the basis of our previous studies,
we investigate structures of $^{10}_{\Lambda}$Be ($\alpha \alpha n\Lambda$)
and $^{10}_{\Lambda}$B ($\alpha \alpha p\Lambda$) and properties of
the underlying $\Lambda N$ interaction.
These $\Lambda$ hypernuclei have provided us many interesting insights so far.
For example, aiming to study $\Lambda N$ spin-dependent interactions,
the high-resolution $\gamma$-ray experiment was performed to
measure the splitting  of the $1^-$-$2^-$ levels of $^{10}_{\Lambda}$B
in BNL-E930 \cite{BNL930}.
However, they  observed no $\gamma$ transition between the
ground state doublet:
this suggests that the
$1^-$-$2^-$ energy splitting in $^{10}_{\Lambda}$B
is less than 100 keV,
or the ground state of this hypernucleus is  a $2^-$ state.

In order to explain the energy splitting in this hypernucleus,
the shell model calculation including $\Lambda N-\Sigma N$
coupling explicitly was performed by Millener \cite{Millener06},
where observed spectra of $p$-shell hypernuclei were reproduced
systematically with the five parameters giving $p_N s_\Lambda$
two-body matrix elements. When this analysis was applied
straightforwardly to $^{10}_{\Lambda}$B, they obtained the ground
$1^-$ state and the $1^-$-$2^-$ splitting energy of 120 keV
\cite{Millener10}.
This splitting is slightly larger than the above limitation energy
100 keV to observe the M1 transition from $2^-$ to $1^-$ state.
They showed also that another interaction set could give rise to
the far smaller value 34 keV\cite{Millener10}
and they mentioned the
$\Lambda \Sigma$ coupling
interaction in this case was unrealistic.
Thus, it is not so simple to reproduce the splitting energy
less than 100 keV in the shell model analysis.
It is very important to investigate the level structures
of $^{10}_{\Lambda}$Be and $^{10}_{\Lambda}$B
within the framework of $\alpha \alpha N \Lambda$
four-body cluster model.
It is reasonable to employ
$\alpha \alpha N \Lambda$ four-body model,
since the core nuclei $^9$B and $^9$Be
are well described by using $\alpha \alpha N$
three-body cluster model,
and, therefore, it should be
possible to model
the structure change of $^9$B and $^9$Be
due to the addition of one $\Lambda$
particle as four-body problem.

Another interesting insight is related to
the charge symmetry breaking (CSB) components in
the $\Lambda N$ interaction.
It is considered that the most reliable
evidence for  CSB  appears in the
$\Lambda$ binding energies $B_\Lambda$ of the $A=4$ members with
$T=1/2$ ($^4_\Lambda$He and $^4_\Lambda$H). Then, the CSB effects
are attributed to the differences
$\Delta_{CSB}=B_\Lambda(^4_\Lambda$He$)-B_\Lambda(^4_\Lambda$H),
the experimental values of which are $0.35\pm0.06$ MeV and
$0.24\pm0.06$ MeV for the ground ($0^+$) and excited ($1^+$)
states, respectively.

The pioneering idea
for the origin of the CSB interaction
was given in Ref.~\citen{DH}, where  $\Lambda$-$\Sigma^0$
mixing leads to an OPEP-type CSB interaction.
This type of meson-theoretical CSB model was shown
yield a  $\Delta_{CSB}$ value for the $0^+$ state in
$^4_\Lambda$He and $^4_\Lambda$H more or less consistent with
to the experimental value. Such interactions, however,
could not reproduce the $\Delta_{CSB}$ value for
the $1^+$ state~\cite{Coon,Usmani}.

The CSB effect is generated also by treating
the masses of $\Sigma^{\pm,0}$ explicitly in
$(NNN\Lambda)$+$(NNN\Sigma)$ coupled four-body calculations
of $^4_\Lambda$He and $^4_\Lambda$H.
In  modern YN interactions such as
the NSC models~\cite{NSC89}~\cite{NSC97},
both elements of the $\Lambda$-$\Sigma^0$ mixing and
the mass difference of $\Sigma^{\pm,0}$ are taken into account.
The exact four-body calculations for $^4_\Lambda$He and
$^4_\Lambda$H were performed using NSC89/97e models
in Ref.~\citen{NKG}. It was shown here that the CSB effect
was brought about dominantly by the $\Sigma^{\pm,0}$
mass-difference effect.
The calculated value of $\Delta_{CSB}$ in the $0^+$ state
was rather smaller than (in good agreement with)
the experimental value for NSC97e (NSC89).
In case of the $1^+$ states, the $\Delta_{CSB}$ value for NSC97e
had the opposite sign from the observed value, and
there appeared no bound state for NSC89.

Thus, the origin of the CSB effect in $^4_\Lambda$He
and $^4_\Lambda$H is still an open question.

As an another approach, phenomenological
central CSB interactions were introduced
in Refs.~\citen{DHT}~\citen{Usmani} so as to
reproduce the $\Delta_{CSB}$ values
apart from the origin of the CSB effect.
Our present work is along this line:
we introduce a phenomenological central CSB interaction
so as to reproduce the $\Delta_{CSB}$
values of $^4_{\Lambda}$H and
$^4_{\Lambda}$He, and use this CSB interaction
in order to investigate the CSB effects in heavier systems.
There exist mirror hypernuclei in the $p$-shell region
such as the $A=7$, $T=1$ multiplet ($^7_\Lambda$He, $^7_\Lambda$Li$^*$,
$^7_\Lambda$Be), $A=8$, $T=1/2$ multiplet
($^8_\Lambda$Li, $^8_\Lambda$Be), $A=10$, $T=1/2$ multiplet
($^{10}_\Lambda$Be, $^{10}_\Lambda$B), and so on.
Historically, some authors mentioned CSB effects in these $p$-shell
$\Lambda$ hypernuclei \cite{Gal77,Gibson95}.

In the past,  accurate estimates of CSB effects in
the $p$-shell region have been  of limited consideration,
because the Coulomb-energies contribute far more
than the CSB interaction \cite{Gibson95}:
There has been  no microscopic calculation of these
hypernuclei taking account of the CSB interaction.
Recently, in Ref. \citen{Hiyama06},
we studied for the first time the CSB effects in
$^7_{\Lambda}$He, $^7_{\Lambda}$Li and $^7_{\Lambda}$Be within
the $\alpha +\Lambda +N+N$ four-body model,
and those in $^8_{\Lambda}$Li and $^8_{\Lambda}$Be
within the $\alpha +t(^3$He)$+\Lambda$ three-body model
using the phenomenological even-state CSB interaction determined
in $^4_\Lambda$He and $^4_\Lambda$H.
This CSB interaction leads to be
inconsistent with the observed data for
 $^8_{\Lambda}$Li
and $^8_{\Lambda}$Be. Then,
as a trial, we introduced
an  odd-state component of the CSB interaction with
opposite sign to the even-state CSB
so as  to reproduce the observed binding energies of
$^8_{\Lambda}$Li and $^8_{\Lambda}$Be.
It is likely that this odd-state CSB interaction
contributes to binding energies of
$A=7$ and 10 $\Lambda$ hypernuclei
as long as we use the even-state CSB interaction to
reproduce the observed binding energies of $A=4$
hypernuclei.
Recently,  a  new experimental data
for  $^7_{\Lambda}$He by $(e,e'K^+)$ were
reported at Thomas Jefferson National
Accelerator Facility (JLab) \cite{Hashimoto2011}.

In this work,
we study $A=10$ hypernuclei within the framework of
an $\alpha +\alpha +N +\Lambda$ four-body model
so as to take account of the full correlations among
all the constituent sub-units.
Two-body interactions
among constituent units are
chosen so as to reproduce all the existing
binding energies of the sub-systems ($\alpha N, \alpha \alpha \Lambda,
\alpha \Lambda$, and so on).
The analysis is performed systematically for ground and  excited states
of the $\alpha \alpha N \Lambda$ systems
with no more adjustable parameters in this stage,
so that these predictions offer important guidance for the
interpretation of the upcoming hypernucleus experiments such as
the $^{10}$B$(e,e'K^+)$ $^{10}_{\Lambda}$Be reaction at JLab.
The CSB effects in binding energies of $^{10}_{\Lambda}$B
and $^{10}_{\Lambda}$Be are investigated in our four-body model
using the even-state CSB interaction determined
in $^4_\Lambda$He and $^4_\Lambda$H.
Furthermore, we introduce trially an odd-state CSB interaction
with opposite sign to the  even-state CSB part
so as to reproduce data of $A=7$ hypernuclei,
and apply it to the present A=10 systems.

In Sec. 2, the microscopic $\alpha \alpha \Lambda N$
calculation method is described.
In Sec.3, the interactions are explained.
The calculated results and the discussion are presented in Sec.4.
Sec. 5 is devoted to a discussion of charge symmetry
breaking effects obtained for the $A=10$ systems.
The summary is given in Sec. 6.

\section{Four-body cluster model and method}

In this work, the hypernuclei,
$^{10}_{\Lambda}$B and $^{10}_{\Lambda}$Be
are considered  to be
composed of two $\alpha$ clusters, a $\Lambda$
particle, and a nucleon.
The core $\alpha$ clusters are
considered to be an inert core and to have
the $(0s)^4$ configuration, $\Psi (\alpha)$.
The Pauli principle between the valence nucleon and the 
nucleons in $\alpha$ clusters 
is taken into account by the orthogonality condition model 
(OCM) \cite{Saito69}, 
as the valence nucleon's wave function should be orthogonal
to  nucleons in the $\alpha$ cluster.

\begin{figure}[htb]
\centerline{\includegraphics[width=8.5 cm,height=8.5 cm]
      {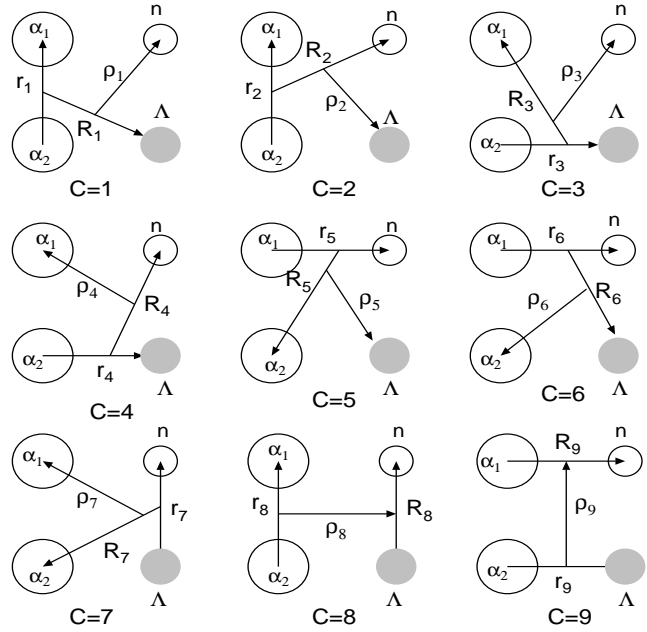}}
\caption{Jacobi coordinates for all
the rearrangement channels ($c=1 \sim 9$)
of the $\alpha +\alpha +\Lambda +N$ four-body system.
Two $\alpha$ clusters are to be symmetrized.}
\label{fig:jacobi10}
\end{figure}

Nine sets of  Jacobian coordinates for the
four-body system of $^{10}_{\Lambda}$B and $^{10}_{\Lambda}$Be
are illustrated in Fig. \ref{fig:jacobi10},
in which we further take into account the
symmetrization between the two $\alpha$s.

%
%
The total Hamiltonian and the   
Schr\"{o}dinger equation
are given by 
\begin{equation}
 ( H - E ) \, \Psi_{JM}(^{\:10}_{\Lambda}{\rm Z})  = 0 \ , 
\label{eq:hamiltonian-10}
\end{equation}

\begin{equation}
 H=T+\sum_{a,b}V_{a b}
      +V_{\rm Pauli} \ ,
\label{eq:hamiltonian-10-1}
\end{equation}
where $T$ is the kinetic-energy operator and
$V_{ab}$ is the interaction between 
constituent particles $a$ and $b$.
The  OCM projection
operator $V_{\rm Pauli}$ will be given below.
The total wavefunction
is described as a sum of amplitudes
of the rearrangement channels $(c=1 \sim$ and 9)
of Fig.~\ref{fig:jacobi10} in the $LS$ coupling
scheme:
\begin{eqnarray}
      \Psi_{JM}\!\!&(&\!\!^{\: 10}_{\Lambda}{\rm Z})
       =  \sum_{c=1}^{9}
      \sum_{n,N,\nu}  \sum_{l,L,\lambda}
       \sum_{s,I,K}
       C^{(c)}_{nlNL\nu\lambda S IK} \nonumber  \\
      &  \times & {\cal S}_\alpha
      \Big[
       \Phi(\alpha_1)\Phi(\alpha_2) 
     \big[ \chi_{\frac{1}{2}}(n)
       \chi_{\frac{1}{2}}(\Lambda)
       \big]_s   \nonumber \\
  & \times  &    \big[ \big[ \phi^{(c)}_{nl}({\bf r}_c)
         \psi^{(c)}_{NL}({\bf R}_c)\big]_I
        \xi^{(c)}_{\nu\lambda} (\mbox{\boldmath $\rho$}_c)
        \big]_{K}  \Big]_{JM} \; .
\label{eq:wavefunction}
\end{eqnarray}
Here the operator $\cal{S}_\alpha$ stands for
symmetrization between the two $\alpha$ clusters.
$\chi_{\frac{1}{2}}(\Lambda)$ and
$\chi_{\frac{1}{2}}(N)$ are
 the spin functions of the $\Lambda$ and nucleon,
 respectively.

Following the Gaussian Expansion Method (GEM)
\cite{Kami88,Kame89,Hiyama03},
we take the functional forms of
$\phi_{nlm}({\bf r})$,
$\psi_{NLM}({\bf R})$ and
$\xi^{(c)}_{\nu\lambda\mu} (\mbox{\boldmath $\rho$}_c)$ as
\begin{eqnarray}
      \phi_{nlm}({\bf r})
      &=&
      r^l \, e^{-(r/r_n)^2}
       Y_{lm}({\widehat {\bf r}})  \;  ,
 \nonumber \\
      \psi_{NLM}({\bf R})
      &=&
       R^L \, e^{-(R/R_N)^2}
       Y_{LM}({\widehat {\bf R}})  \;  ,
 \nonumber \\
      \xi_{\nu\lambda\mu}(\mbox{\boldmath $\rho$})
      &=&
       \rho^\lambda \, e^{-(\rho/\rho_\nu)^2}
       Y_{\lambda\mu}({\widehat {\rhovec}})  \; ,
\end{eqnarray}
where the Gaussian range parameters are chosen 
according to geometrical progressions:
\begin{eqnarray}
      r_n
      &=&
      r_1 a^{n-1} \qquad \enspace
      (n=1 - n_{\rm max}) \; ,
\nonumber\\
      R_N
      &=&
      R_1 A^{N-1} \quad
     (N \! =1 - N_{\rm max}) \; ,  
\nonumber\\
      \rho_\nu
      &=&
      \rho_1 \alpha^{\nu-1} \qquad
     (\nu \! =1 - \nu_{\rm max}) \; .  
\end{eqnarray}
  The eigenenergy $E$  in Eq.(2.1)
and the coefficients $C$ in Eq.(2.3) 
are  determined by the Rayleigh-Ritz variational method.

The Pauli principle between nucleons belonging
to two $\alpha$  clusters
is taken into account by the orthogonality condition
model (OCM) \cite{Saito69}.
The OCM projection operator $V_{\rm Pauli}$
appearing in Eq. (2.2)
is represented by
\begin{equation}
V_{\rm Pauli}=\lim_{\gamma\to\infty} \ \gamma \
\sum_f
|\phi_f({\bf r}_{\alpha x})
\rangle \langle \phi_f({\bf r}'_{\alpha x})| \:,
\label{eq:ocm}
\end {equation}
which rules out the amplitude of the
Pauli-forbidden $\alpha -\alpha$ and 
$\alpha$-$n$ relative
states $\phi_f({\bf r}_{\alpha x})$
from the four-body total wavefunction \cite{Kukulin95}.
The forbidden states are
$f={0S,1S,0D}$ for $x=\alpha$ and $f=0S$ for $x=n$, respectively.
The Gaussian range parameter $b$ of the
single-particle $0s$ orbit in the
$\alpha$ cluster $(0s)^4$ is taken to be
$b=1.358$ fm so as to reproduce the
size of the $\alpha$ cluster.
In the actual calculations, the strength $\gamma$
for $V_{\rm Pauli}$ is taken to be
$10^4$ MeV, which is large enough to push
the unphysical forbidden state to
the very high energy region, while keeping the
physical states unchanged.

\section{Interactions}

\label{interaction}

\subsection{Charge symmetry parts}

For $V_{N \alpha}$, we employ the effective potential
proposed in Ref.\citen{Kanada79}, which is designed so as to reproduce
well  low-energy scattering phase
shifts of the $\alpha N$ system.
The Pauli principle between nucleons belonging to the
$\alpha$ and the valence nucleon is taken into account
by the orthogonality condition model (OCM) \cite{Saito69} as mentioned
before.

For $V_{\Lambda N}$, we employ the same as used in the
structure calculations of $A=7$ hypernuclei in
Refs.\citen{Hiyama06,Hiyama09}.
Namely, this is an effective single-channel
interaction simulating the basic features of the
Nijmegen model NSC97f~\cite{NSC97}, where
the $\Lambda N$-$\Sigma N$ coupling effects are renormalized
into $\Lambda N$-$\Lambda N$ parts:
We use three-range Gaussian potentials designed to reproduce
the $\Lambda N$ scattering phase shifts calculated
from  NSC97f, with
their second-range strengths in the $^3E$ and $^1E$ states
adjusted so that the calculated energies of
the $0^+$-$1^+$ doublet state in the $NNN\Lambda$
four-body system chosen to  reproduce the observed
splittings of $^4_\Lambda$H.
Furthermore, the spin-spin parts in the odd states are tuned
to yield the experimental values of
the splitting energies of $^7_\Lambda$Li.
The symmetric LS (SLS) and anti-symmetric LS (ALS)
parts in $V_{\Lambda N}$ are chosen so as to
be consistent with the $^9_\Lambda$Be data:
The SLS and ALS parts derived from NSC97f with the G-matrix
procedure are represented in the two-range form, and then
the ALS part is enhanced so as to reproduce
the measured $5/2^+$-$3/2^+$ splitting energy in the
$2\alpha + \Lambda$ cluster model~\cite{Hiyama00}.

The interaction $V_{\alpha \Lambda}$ is obtained by folding the
$\Lambda N$ G-matrix interaction derived from
the Nijmegen model F(NF)~\cite{NDF}
with the density of the $\alpha$ cluster~\cite{Hiyama97}, its strength
being adjusted so as to reproduce the experimental value of
$B_\Lambda(^5_\Lambda$He).
Furthermore, we use $\alpha \Lambda$ SLS and ALS terms
which are obtained by folding the same
$\Lambda N$ SLS and ALS parts as 
mentioned before.

For $V_{\alpha \alpha}$,
we employ the potential that has
been used often in the OCM-based cluster-model
study of light nuclei \cite{Hasegawa71}.
The potential reproduces reasonably well 
the  low-energy scattering phase shifts
of the $\alpha \alpha$ system.
The Coulomb potentials are constructed by folding the
$p$-$p$ Coulomb force with the
proton densities of all the participating clusters.
Since the use of the present $\alpha \alpha$ and
$\alpha n$ interactions does not precisely
reproduce the energies of the
low-lying states of $^9$Be as measured from
the $\alpha \alpha n$ threshold,
we introduce an additional phenomenological
$\alpha \alpha n$ three-body force 
so as to fit the observed energies of the
$3/2^-_1$ ground state and
$5/2^-_1$, $1/2^-_1$ and $1/2^+_1$ excited states in
$^9$Be.
The parameters of this $\alpha \alpha n$ three-body force
are listed in Ref.\citen{Hiyama2011}.
This $V_{\alpha \alpha}$ potential is applied to
the three-body calculation of the $\alpha \alpha p$ system, and
the energy of the ground state reproduces
the observed data well.

\subsection{Charge symmetry breaking interaction}

\label{sec:CSB}

It is beyond the  scope in this work to explore the origin of
the CSB interaction. We employ  here the 
following phenomenological CSB interaction
with  one-range Gaussian form :
\begin{eqnarray}
&& V_{\Lambda N}^{\rm CSB}(r)= 
-\frac{\tau_z}{2}\Big[\frac{1+P_r}{2}(v_{0}^{\rm even,CSB} +
\mbox{\boldmath $\sigma$}_\Lambda \cdot
\mbox{\boldmath $\sigma$}_N
v_{\sigma_\Lambda \cdot
\sigma_N}^{\rm even, CSB})
\, e\:^{-\beta_{\rm even}\: r^2}\   \nonumber
\\ 
&+& \frac{1-P_r}{2}(v_{0}^{\rm odd,CSB} +
\mbox{\boldmath $\sigma$}_\Lambda \cdot
\mbox{\boldmath $\sigma$}_N
v_{\sigma_\Lambda \cdot
\sigma_N}^{\rm odd,CSB})
\, e \:^{-\beta_{\rm odd} \:r^2}\ \Big]  ,
\end{eqnarray}
which includes spin-independent and spin-spin parts.
The range parameter, $\beta_{\rm even}$ is taken to be 1.0 fm$^{-2}$.
The parameters $v_0^{\rm even}$ and $v_{\sigma \sigma}^{\rm even}$
are determined phenomenologically so as to reproduce the values of
$\Delta_{CSB}$ derived from the $\Lambda$ binding energies of
the $0^+$ and $1^+$ states in the
four-body calculation of $^4_\Lambda$H ($^4_\Lambda$He).
Then, we obtain $v_0^{\rm even,CSB}=8.0$ MeV and
$v_{\sigma \sigma}^{\rm even,CSB}$=0.7 MeV.

In order to extract the information about the odd-state part of CSB,
it is necessary to study iso-multiplet hypernuclei in the $p$-shell region.
A suitable system for such a study is $^7_{\Lambda}$He,
in which the core nucleus $^6$He is in a bound state.
The JLab E01-011 experiment 
measured $^7$Li $(e,e'K^+)^7_{\Lambda}$He reaction 
and  reported  the binding energy of
$^7_{\Lambda}$He ground state to be
$5.68 \pm 0.03 \pm 0.25$ MeV for the first time
\cite{Hashimoto2011,Nakamura2012}.
The present experimental data has a large systematic
error which is the same order to
the discussing CSB effect.
They measured the same reaction with the
improved calibration in the JLab E05-115 experiment 
\cite{E05-115} and 
more accurate result will be obtained 
in near future. Before the final experimental
result of result of $^7_{\Lambda}$He is
obtained, we will use our calculated
binding energy of $^7_{\Lambda}$He,
$B_{\Lambda}=5.36$ MeV which
locates around the limit of the 
current experimental error,
to tune the strength and range of the
odd-state.
The range parameter,  $\beta_{\rm odd}$
is taken to be $1.5$ fm.
The strengths , $v_0^{\rm odd,CSB}$
, $v_{\sigma \sigma}^{\rm odd, CSB}$
are taken to be 16.0 MeV and 
0.7 MeV, respectively.
Using these potential parameters,
the $\Lambda$-separation energy of  the mirror 
$\Lambda$ hypernucleus, $^7_{\Lambda}$Be 
is 5.27 MeV, which reproduces the observed data, too.

\begin{figure}[htb]
\begin{center}
\centerline{\includegraphics[width=12 cm,height=12.0 cm]
{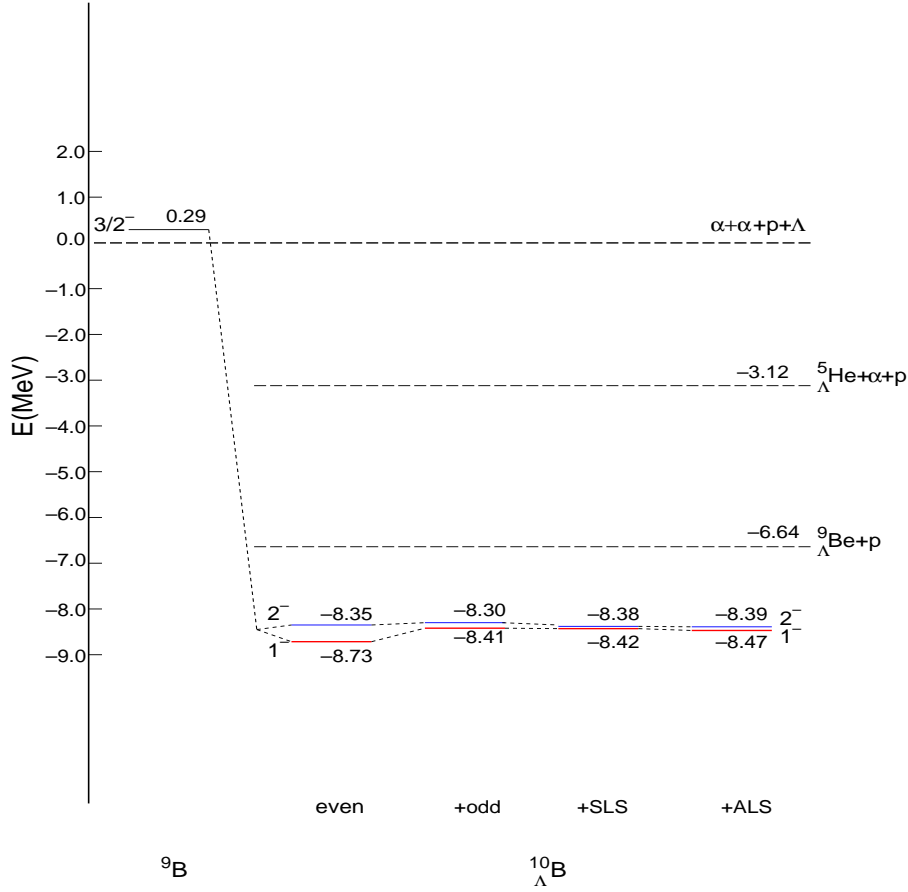}}
\label{fig:b10l}
\caption{(color online).
Calculated energy levels of $^9$B and $^{10}_{\Lambda}$B.
The charge symmetry breaking potential is not included in
 $^{10}_{\Lambda}$B. The level energies are measured 
 with respect to $\alpha +\alpha +\Lambda +p$
particle breakup threshold.}
\end{center}
\end{figure}

\section{Results}

\subsection{spin doublet states of $A=10$ hypernuclei}

First, let us describe the level structures of $^{10}_{\Lambda}$B
and $^{10}_{\Lambda}$Be obtained
with the $\alpha +\Lambda +N+N$ four-body model,
when the CSB interaction is not included.
Calculations are performed for four-body
bound states in these $\Lambda$ hypernuclei.

In Figs. 2 and 3 and in Table I, we show the level structures
of $^{10}_{\Lambda}$B and $^{10}_{\Lambda}$Be. 
In each figure, hypernuclear levels are shown
in four columns in order to demonstrate
separately the effects of the even-state and odd-state
$\Lambda N$ interactions, and also the SLS and ALS
interactions.
Even when the CSB interactions are switched on,
their small contributions do not alter the features of 
these figures.
Table I gives calculated values of $\Lambda$ binding
energies and root mean square (r.m.s.) distances of subsystems
in $^{10}_{\Lambda}$B and $^{10}_{\Lambda}$Be.

It is considered that the $1^-$-$2^-_1$ spin-doublet
states in $^{10}_{\Lambda}$B and $^{10}_{\Lambda}$Be,
and also the $2^-_2$-$3^-$ and $0^+$-$1^+$
spin-doublet states in 
$^{10}_{\Lambda}$Be, give useful information about
the underlying spin-dependence of the $\Lambda N$
interaction.
It should be noted that the $\Lambda N$ interaction used
in the present calculations is identical to 
the one used 
in our previous analyses of the $T=0$ spin-doublet state
of $^7_{\Lambda}$Li \cite{Hiyama06} and
the $3/2^+$-$5/2^+$ spin-doublet states
of $^7_{\Lambda}$He and $^7_{\Lambda}$Li with $T=1$
\cite{Hiyama09}.
That is no  additional parameter for adjusting to
the experimental data is used in the present calculations.

%
As shown in Fig. 2, we see that the resultant energy splitting of
the $1^-$-$2^-_1$ states in
$^{10}_{\Lambda}$B
is  0.08 MeV, with
combined contributions from the
spin-spin, SLS and ALS interactions.
For the study of the $\Lambda N$ spin-dependent interaction,
in BNL-E930, they tried to measure the $1^-$-$2^-_1$
spin-doublet states in $^{10}_{\Lambda}$B
using the $^{10}$B($K^-,\pi^- \gamma)$ reaction.
However, the $M1$ transition between the ground-state
doublet members ($2^-_1 \rightarrow 1^-)$ was
not observed. 
%
This measurement suggests the following two possibilities:
The energy splitting between the $1^-$ and $2^-_1$ states is less than 100 keV
and the $\gamma$ ray cannot be observed, since the $\gamma$-ray detection
efficiency drops rapidly below 100 keV.
 The other suggestion is that  the $2^-_1$ state,
 which is dominantly  
 produced by the
 $(K^-, \pi^- \gamma)$ reaction,
is  
 the ground state.
 Our result supports the former.

Next, let us see in more detail how the
$\Lambda N$ spin-spin interactions contribute to the
$1^-$-$2^-$ doublets in $^{10}_{\Lambda}$B.
The $1^-$ state is composed of $[K=1 (N\Lambda)_{s=0,1}]_{J=1^-}$
and $[K=2 (N\Lambda)_{s=1}]_{J=1^-}$,
where the $K$ is the angular momentum and $s$ is
the spin of $N-\Lambda$ described in Eq. (2.3).
Among these three components,
$[K=1 (N\Lambda)_{s=0,1}]_{J=1^-}$ components are comparable with each other.
On the other hand,
the $2^-_1$ state is composed of 
$[K=1 (N\Lambda)_{s=1}]_{J=2^-}$
and $[K=2 (N\Lambda)_{s=0, 1}]_{J=2^-_1}$ components,
where
the $[K=1 (N\Lambda)_{s=1}]_{J=2^-_1}$
and $[K=2 (N\Lambda)_{s=1}]_{J=2^-_1}$ components
are larger than the other one.
The  $V_{\Lambda N}({^1E})$ is
more attractive than the $V_{\Lambda N}({^3E})$,
when they are adjusted so as to reproduce 
the $0^+$-$1^+$ splitting energy 
in $^4_\Lambda$H ($^4_\Lambda$He).
In $^{10}_{\Lambda}$B,
this even-state interaction makes
the $1^-$ state lower than $2^-_1$ state.
The value obtained for the splitting energy is 0.38 MeV.
This value is far larger than the above limitation of 100 keV
suggested by the no  $\gamma$-ray observation in BNL-E930.
On the other hand,
the calculated value of $B_{\Lambda}$ in the ground state 
is 9.02 MeV,
which is consistent with  
 the experimental value of 
$8.89 \pm 0.12$ MeV within the error bar.

Next, when the odd-state
interaction is switched on,
the energy splitting is reduced to 0.11 MeV (see ''+odd'' column).
The reason for this reduction is
because $V_{\Lambda N}({^1O})$ is
more repulsive than $V_{\Lambda N}({^3O})$
as indicated in our analysis for the $1/2^+$-$3/2^+$
spin-doublet state in $^7_\Lambda$Li.
Then, in $^{10}_{\Lambda}$B,
the $1^-$ state including dominantly the $\Lambda N$
spin-singlet component is pushed up
more than the $2^-_1$ state.

Moreover, we study the effects of the SLS and ALS interactions
on $1^-$ and $2^-_1$ doublet states.
As shown in Fig.2, the SLS works attractively for the
$2^-_1$ state because the contribution of the
$\Lambda N$ spin-triplet state is
dominant in this state,
while its contribution is very small to the $1^-$ state
which is dominated by the spin-singlet component. Thus, 
the $1^-$-$2^-_1$ splitting is found to be reduced by the SLS.

On the other hand,
the ALS works significantly in the $1^-$ state,
because the ALS acts
between the spin$=0$ and 1 $\Lambda N$ components
and both of them are included in the $1^-$ state.
However, the ALS contribution is not significant in the
$2^-_1$ state, 
because this state is dominated by the spin =1 $\Lambda N$
component.

As a result of including both the spin-spin
and spin-orbit terms,
the energy splitting of the $1^-$-$2^-_1$ states
of $^{10}_{\Lambda}$B leads to be 0.08 MeV.
Then, we obtain the calculated value
$B_{\Lambda}(^{10}_{\Lambda}$B)=8.76 MeV
which does not differ significantly from  the
experimental value, $8.89 \pm 0.12$ MeV.

We can see the same tendency in
$^{10}_{\Lambda}$Be
and the resultant energy splitting is
0.08 MeV, which is the same as that of $^{10}_{\Lambda}$B, 
as shown in Fig.3.
The calculated $B_{\Lambda}$ value of the ground state is 8.94 MeV.
As in the $^{10}_{\Lambda}$B case, this value is rather close to
the experimental value 
$B_{\Lambda}(^{10}_{\Lambda}$Be)=$9.11 \pm 0.22$ MeV.
A more detailed discussion of the
binding energies of $^{10}_{\Lambda}$B and
$^{10}_{\Lambda}$Be
with/without CSB interaction will appear
in the next session. 
%
Next, let us discuss one more spin-doublet state,  
$2^-_2$-$3^-$,  in $^{10}_{\Lambda}$Be.
The dominant component of the $2^-_2$ ($3^-$) state 
is $[K=2 (N\Lambda)_{s=0}]_{J=2^-_2}$ 
($[K=2 (N\Lambda)_{s=1}]_{J=3^-}$).
Then, with the use of our even state interaction,
the $2^-_2$ state is lower than $3^-$ state and
the energy splitting is 0.43 MeV.
When the odd state interaction is included in 
calculations of $2^-_2$ and $3^-$ states,
the energy of the $2^-_2$ state is pushed up more than that 
of the $3^-$ state due to the repulsive contribution of the
$V_{\Lambda N}^{(^1O)}$ component
and energy splitting is 0.12 MeV.
When the SLS interaction is added to the calculations of these states,
the SLS contributes dominantly to the $3^-$ state.
Finally, the repulsive ALS interaction,
having the opposite sign of the SLS, contributes mainly
to the $2^-_2$ state including both of spin-singlet and 
spin-triplet states. As a result, we have
0.05 MeV for the $3^-$-$2^-_2$ doublet splitting.

Furthermore, the above the $3^-$ and $2^-_2$ states,
we have $0^+$ and $1^+$ states as bound states which
is composed of $^9{\rm Be}(1/2^+)+\Lambda(0s_{1/2})$.
In the core nucleus, $^9$Be, 
the $1/2^+$ state is observed as the first excited state
and is lower than the $5/2^-$ and $1/2^-$
excited states, despite the last neutron in this
$1/2^+$ state presumably occupying the
$1s_{1/2}$ orbit in the
simple shell model configuration, whereas
the next two excited states
with negative parity would have $1p$-shell
configurations.
It is interesting to see
that the order of the $3^-$, $2^-_2$, $0^+$ and
$1^+$ states is reversed
from $^9$Be to $^{10}_{\Lambda}$Be.
The $5/2^-$ state is composed of
$^8{\rm Be}(2^+)+n(p_{1/2},p{3/2})$
and then there is a centrifugal barrier between $\alpha \alpha$
and a valence neutron, while the
$1/2^+$ state does not have any barrier.
Thus it is considered that the $5/2^-$ state
is more compact than the $1/2^+$ state.
When a $\Lambda$ particle adds into these
states, we see the energy gain is larger than
in the compactly coupled state ($5/2^-$) than
in the loosely coupled state $(1/2^+$).
It should be noted that the
same type of theoretical prediction was reported
in our early work \cite{Hiyama00} for
the $\alpha \alpha \alpha \Lambda$ four-body model
of $^{13}_{\Lambda}$C, where the
$\Lambda$ particle is added to the
compact bound state $(3^-_1$) and to
the loosely bound state $(0^+_2$) in
$^{12}$C.

Let us discuss about energy splitting of this positive
parity states.
The dominant component of the
$0^+$ $(1^+$) is $[K=0 (N\Lambda)_{s=0}]_{0^+}$
($[K=0 (N\Lambda)_{s=1}]_{1^+}$).
Then using the even state spin-spin interaction,
the $0^+$ state is lower than the $1^+$ state
and energy splitting is 0.57 MeV.
When the odd state spin-spin interaction
is employed, the energy of the $0^+$
is pushed up more than that of the
$1^+$ state due to the repulsive contribution
of the $V_{\Lambda N}^{(^1O)}$ component
and the
energy splitting is  $0.26$MeV.
Since these two states are composed of
$^9{\rm Be}(1/2^+)+\Lambda(0s_{1/2})$ as mentioned before,
then the relative angular momenta between
composed particles are almost $s$-wave,
then  spin-orbit contribution for these doublets
is very small.
As shown in Fig.3, we see that the contributions of
SLS and ALS for these doublets are small. Thus,
we have 0.2 MeV finally for this positive parity doublet.

In Fig.3, we found that the energy splittings 
of the negative parity doublets are
less than 0.1 MeV, while that of 
the positive parity state is much larger.
The reason is as follows:
The $\alpha N$ spin-orbit interaction makes
$2^-$ state lower than the $1^-$ state.
On the other hand, the $\Lambda N$ spin-spin interaction
makes $1^-$ state lower than the $2^-$ state.
Due to the cancellation between
$\alpha N$ spin-orbit interaction and
$\Lambda N$ spin-spin interaction,
we have less than 0.1 MeV splitting energy.
In order to investigate the efect of the $\alpha N$ spin-orbit 
interaction
for the $1^-$-$2^-$ doublet state,
as a trial, we turn off 
the $\alpha N$ spin-orbit term.
In this case, we use $\Lambda N$ even and odd-state
spin-spin forces.
Then, the energy splitting is obtained to be 0.27 MeV
owing to the even- and odd-state spin-spin forces.
This value is
similar with one of $0^+$-$1^+$ state.
Thus, we find that $\alpha N$ spin-orbit force
give a contribution to the energy splitting of
the ground state doublet.

On the other hand, in the case of
$0^+$ and $1^+$ states,
the $\alpha N$ spin-orbit contribution to
this energy splitting is significantly small,
because composed particles $\alpha$ and $N$ are
in the $s$-state relatively.
As a result, we get the pure contribution  
of the spin-spin $\Lambda N$ interaction for
the energy splitting of $0^+$ and $1^+$ states, 0.2 MeV.

Then, it would be difficult to observe
$\gamma$-ray transitions between the negative
parity spin-doublet partners,
but
it might be possible to observed $\gamma$-ray from
$0^+$ and $1^+$ states.

\begin{figure}[htb]
\begin{center}
\centerline{\includegraphics[width=12 cm,height=12.0 cm]
      {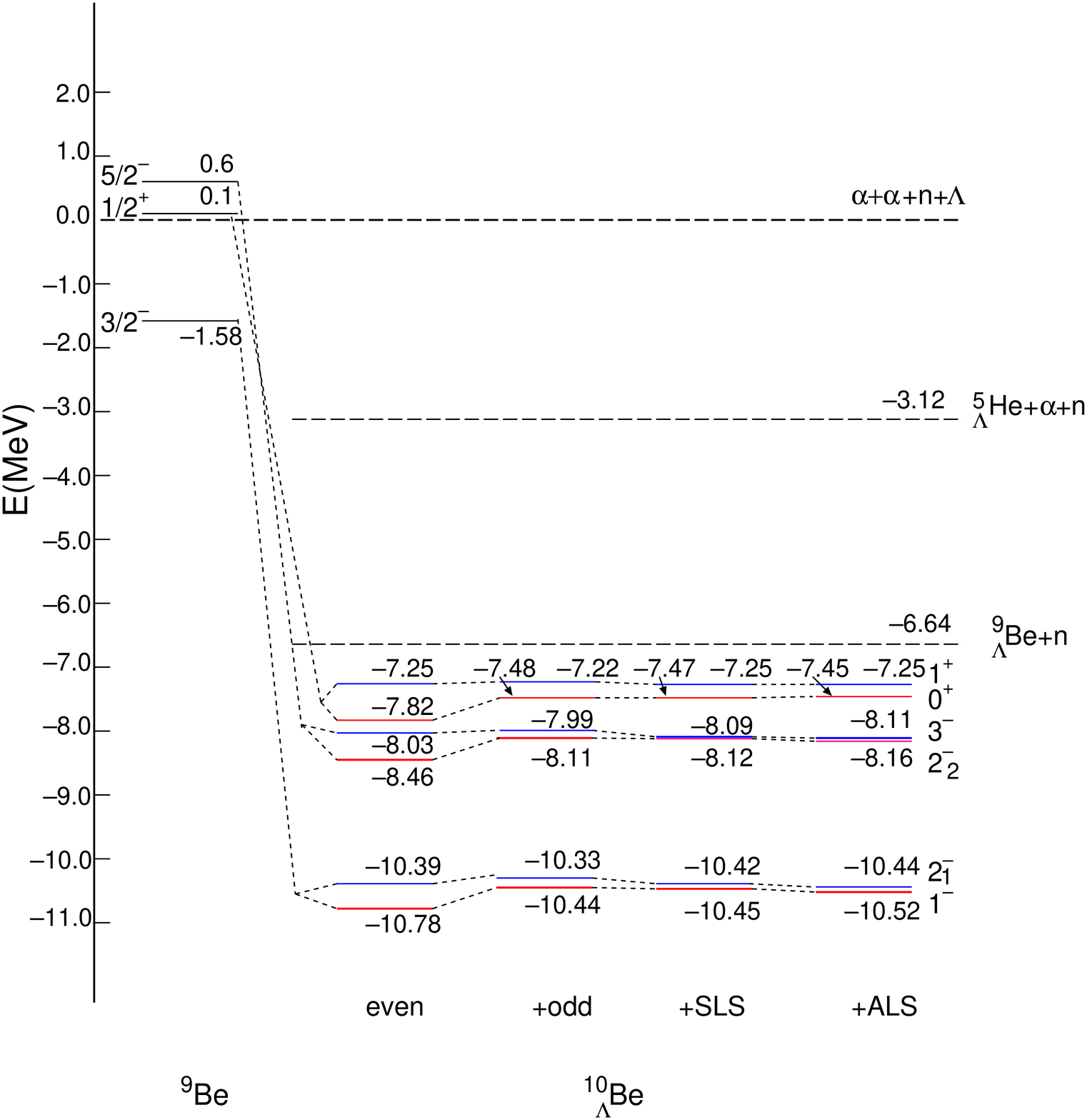}}
\label{fig:be10l}
\caption{(color online).
Calculated energy levels of $^9$Be and $^{10}_{\Lambda}$Be.
The charge symmetry breaking potential is not included in
 $^{10}_{\Lambda}$Be.
 The level energies are measured with respect to
$\alpha +\alpha +\Lambda +n$ particle breakup threshold.}
\end{center}
\end{figure}

\begin{table}
\begin{center}
\caption{
Calculated energies of the low-lying states of
(a) $^{10}_{\Lambda}$B and (b)$^{10}_{\Lambda}$Be
 without the charge symmetry breaking potential,
together
  with those of the corresponding states of $^9$B and $^9$Be
, respectively.
$E$ stands for the total interaction energy among constituent particles.
The energies in the parentheses are measured from the corresponding
lowest particle-decay thresholds $^9_\Lambda$Be$+$ $N$
for $^{10}_{\Lambda}$B and $^{10}_{\Lambda}$Be.
The calculated r.m.s. distances,
$\bar{r}_{\alpha -\alpha}$, $\bar{r}_{\alpha -\Lambda}$
$\bar{r}_{\alpha -n}$, $\bar{r}_{\Lambda -N}$
 are also listed for the bound state.}
\label{tab:rms_A=10}
\begin{tabular}{ccccccccccccc}
\hline \hline  &&&&&&\\[-3mm] && &(a) \\
            &\multicolumn{1}{c}{$^{9}$B($\alpha \alpha p$)} & \hspace{5mm}
   & \multicolumn{2}{c}{$^{10}_{\Lambda}$B$(\alpha \alpha \Lambda p)$} \\
   $J^{\pi}$  &$3/2^-$    & & $1^-$  &$2^-$    \\ \hline %
$E$	(MeV) & $+0.29$   &
& $-8.47$    &$-8.39$   \\
$E^{\rm exp}$(MeV) &$0.28$ &  &    &                \\
&	&         &$(-1.83)$  &$(-1.75)$   \\
$B_{\Lambda}$(MeV) &     &  &8.76  &8.67      \\
$B_{\Lambda}^{\rm exp}$  (MeV)	&  &    &$8.89 \pm 0.12$  &           \\
$\bar{r}_{\alpha-\alpha}$(fm) & & &$3.32$  &$3.30$   \\
$\bar{r}_{\alpha-\Lambda}$(fm) & & &$3.04$  &$3.02$  \\
$\bar{r}_{\alpha-p}$(fm) & & &$3.64$  &$3.64$  \\
$\bar{r}_{\Lambda-p}$(fm) & & &$3.86$  &$3.87$   \\
 && &(b) \\
            &\multicolumn{2}{c}{$^{9}$Be($\alpha \alpha n)$} & \hspace{5mm}
   & \multicolumn{4}{c}{$^{10}_{\Lambda}$Be($\alpha \alpha \Lambda n)$} \\
  $J^{\pi}$  &$3/2^-$   & $5/2^-$     & $1^-$  & $2^-_1$  &$2^-_2$ 
  &$3^-$  &$0^+$  &$1^+$  \\ \hline %
$E$	(MeV) & $-1.58$  &\quad $0.60$  
& $-10.42$   & $-10.38$ &$-8.13$ &$-8.01$   &$-7.45$  &$-7.25$ \\
$E^{\rm exp}$(MeV) &$-1.58$ &\quad $0.85$  &    &   & &            \\
&	&           & $(-3.76)$ & $(-3.74)$ &($-1.47)$ &$(-1.35)$  &$(-0.81)$ 
&$(-0.61$)  \\
$B_{\Lambda}$(MeV) &        & & $8.84$  &$8.80$ &$6.63$ &6.26 &5.87 &5.67  \\
$B_{\Lambda}^{\rm exp}$  (MeV)	   & &   &$9.11 \pm 0.22$ &   &     &&   \\
$\bar{r}_{\alpha-\alpha}$(fm) &$3.68$  &-   &$3.27$ &$3.26$  &$3.29$ &$3.24$ 
&3.78 &3.76
 \\
$\bar{r}_{\alpha-\Lambda}$(fm)   && &$3.02$ &$3.00$  &$3.02$ &$3.00$
&3.31 &3.30  \\ 
$\bar{r}_{\alpha-n}$(fm) &$4.56$  &- &$3.52$ &$3.51$  &$3.56$ &$3.56$ 
&4.95  &5.04   \\ 
$\bar{r}_{\Lambda-n}$(fm)   && &$3.77$ &$3.75$  &$3.85$ &$3.80$  &5.04
&5.15   \\ 
\hline \hline
\end{tabular}
\end{center}
\end{table}

It is interesting to explore the glue-like role
of the $\Lambda$ particle in $^{10}_{\Lambda}$B and $^{10}_{\Lambda}$Be.
Though the ground state of
$^9$B is unbound, the corresponding states ($1^-$, $2^-_1$)
in $\Lambda$ hypernuleus becomes bound by $1.7 \sim 2.0$ MeV
due to the addition of a
$\Lambda$.
On the other hand, the ground state of the core nucleus $^9$Be is
bound by 1.58 MeV with respect to  the
$\alpha +\alpha +n$ three-body threshold.
Owing to an additional $\Lambda$ particle, the
corresponding ground state of $^{10}_{\Lambda}$Be 
become rather deeply bound, by $\sim$ 4 MeV.
Furthermore, the $5/2^-$ resonant state of $^9$Be 
become bound ($3^-$ and $2^-_2$ in $^{10}_{\Lambda}$Be)
by $\sim$ 1.2 MeV due to the presence of
the $\Lambda$ particle.
In addition, when a $\Lambda$ particle
is added to the $1/2^+$ state of $^9$Be,
the $0^+$ and $1^+$ states of $^{10}_{\Lambda}$Be
become weakly bound by less than $1.0$ MeV.
 
In Table I, we list the calculated values of the r.m.s. 
radii between composed particles, $\bar{r}_{\alpha-\alpha}$,
$\bar{r}_{\alpha-\Lambda}$, $\bar{r}_{\alpha-N}$
and  $\bar{r}_{\Lambda-N}$ in our four-body model of
$^{10}_{\Lambda}$B and $^{10}_{\Lambda}$Be. 

From the calculated
rms radii, it is interesting to look at the dynamical change of the
nuclear core $^9$Be, which occurs due to
the addition of a $\Lambda$ particle.
The possibility of nuclear-core shrinkage due
to a addition of $\Lambda$-particle  was originally pointed out
in Ref. \citen{Motoba83} by
using the $\alpha x\Lambda$ three-cluster model
($x=n,p,d,t,^3$He, and $\alpha$) for
$p$-shell $\Lambda$ hypernuclei.
As for the hypernucleus $^7_{\Lambda}$Li,
the prediction of some $20$ \%
shrinkage, in Ref. \citen{Motoba83} and 
in an updated calculation \cite{Hiyama99},
was  confirmed by experiment \cite{Tanida01}.
As shown in Table I, the
rms distance $\bar{r}_{\alpha-\alpha}$ between two $\alpha$ clusters, 
and $\bar{r}_{\alpha-N}$ between $\alpha$ and nucleon, 
are reduced by $12-17$ \% with the addition of
a $\Lambda$ particle.

As shown in Table I, the values of
$\bar{r}_{\alpha-N}$ in these systems
are larger than those of $\bar{r}_{\alpha-\Lambda}$,
indicating that the distributions of
valence nucleons have  longer-ranged tails than those
of the $\Lambda$'s in the respective systems.
Especially, $\bar{r}_{\alpha-N}$ in the $0^+$ and the
$1^+$ states are much larger, around 5 fm
than those of the other states.
Then, it is expected that these positive parity states
have  neutron halos.
 
In order to see the structure of these systems visually,
in Fig. 4, we draw the density distributions of the 
$\Lambda$ (dashed curve) and valence neutrons (solid curve) of
$0^+$ state of $^{10}_{\Lambda}$Be.
For comparison here, also a single-nucleon density
in the $\alpha$ core is shown
by the dotted curve.
You find that we have long-range neutron density
as shown in Fig.4.

%
\begin{figure}[htb]
\centerline{\includegraphics[width=8.5 cm,height=8.0 cm]
      {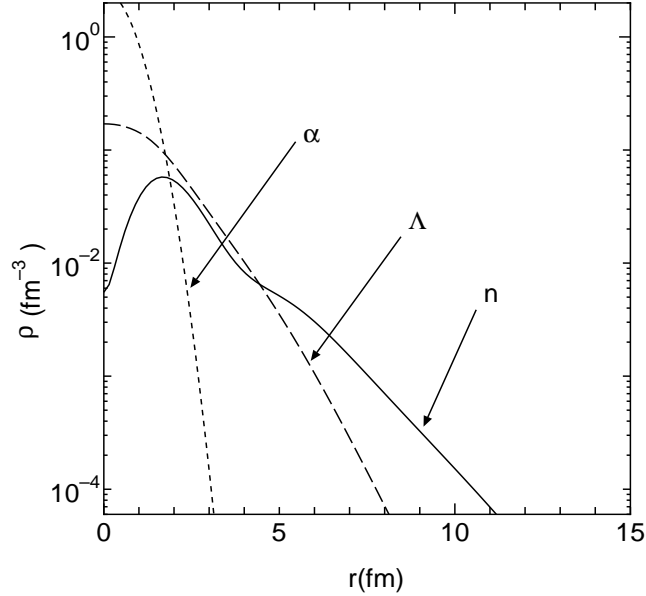}}
\caption{Calculated density distribution of
$\alpha$, a $\Lambda$ and a valence nucleon
for $0^+$ state of $^{10}_{\Lambda}$Be
without a charge symmetry breaking potential.}
\label{fig:halo}
\end{figure}

In addition, we show the
density distributions of 
$1^-$ state of $^{10}_{\Lambda}$Be and $^{10}_{\Lambda}$B
in Fig.5.
In each case,  the density distribution
of the $\Lambda$ has a shorter-ranged tail than
that of a valence nucleon,
but is extended significantly far away from
the $\alpha$ core, which can be thought of as
 three layers of matter 
composed of two $\alpha$ clusters,
a $\Lambda$ , and a nucleon.

\begin{figure}[htb]
\begin{minipage}{0.35\linewidth}
\scalebox{0.40}
{\includegraphics{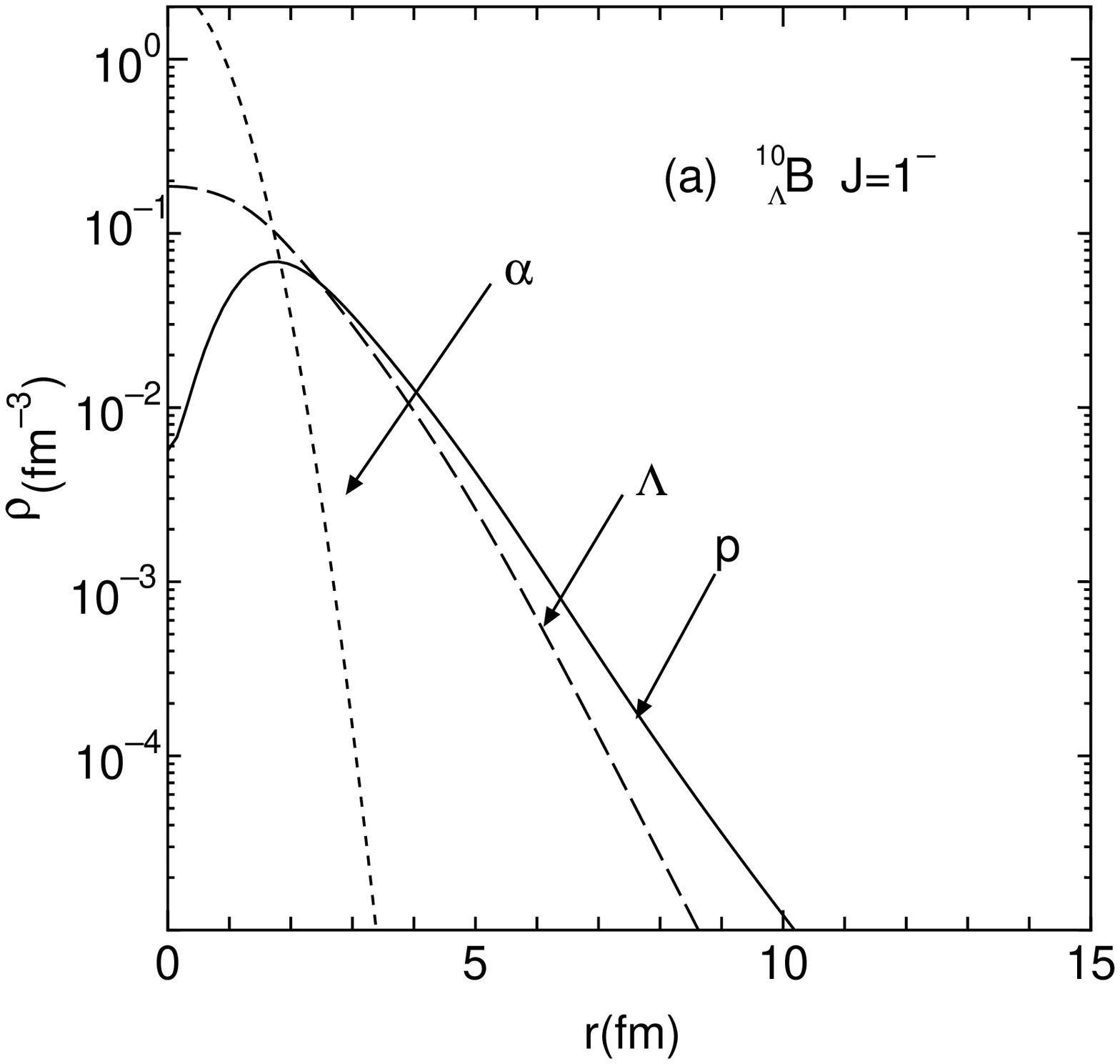}}
\end{minipage}
\hspace{1.8cm}
\begin{minipage}{0.35\linewidth}
\scalebox{0.40}
{\includegraphics{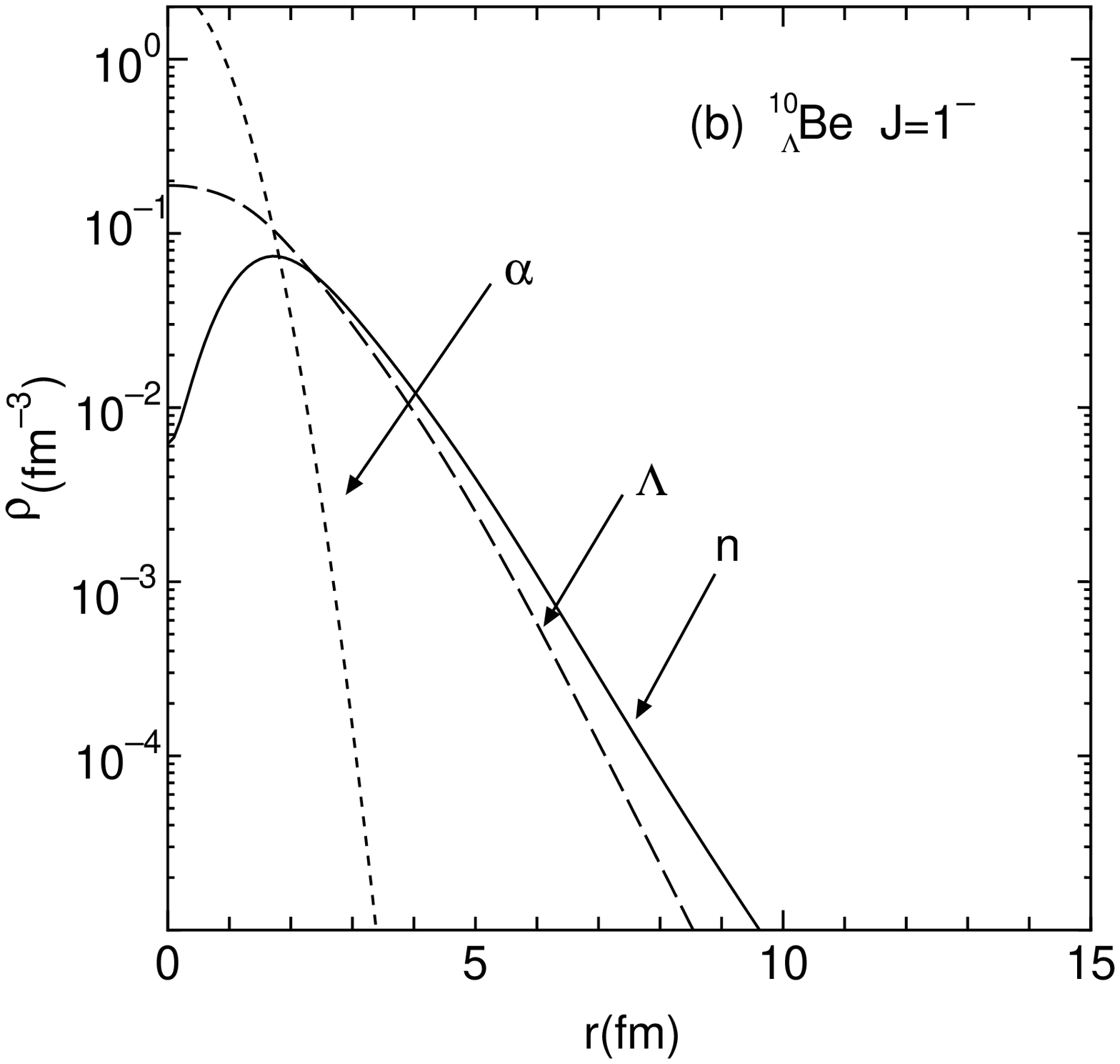}}
\end{minipage}
\caption{Calculated density distribution of
$\alpha$, a $\Lambda$ and a valence nucleon
for (a)$^{10}_{\Lambda}$B and (b)$^{10}_{\Lambda}$Be
without a charge symmetry breaking potential.}
\label{fig:b10den}
\end{figure}

\subsection{Charge Symmetry breaking effects}

Let us focus on the ground states in
$^{10}_{\Lambda}$Be and $^{10}_{\Lambda}$B.
It is likely that the CSB interaction 
affect in binding energies of these isodoublet hypernuclei.

In subsec.3.2, we introduce
the phenomenological CSB potential
with the central-force component only.
The CS part of the two-body $\Lambda N$
interaction is fixed to reproduce
the averaged energy spectra of
$^4_{\Lambda}$H and $^4_{\Lambda}$He, and then 
the even-state part of the CSB interaction is adjusted 
so as to reproduce both the energy levels of 
these hypernuclei. 
In our previous work~\cite{Hiyama09},
this CSB interaction was applied to calculations of
the binding energies of the $A=7$ isotriplet hypernuclei, 
$^7_{\Lambda}$He, $^7_{\Lambda}$Li$(T=1)$,
and $^7_{\Lambda}$Be.
Here, the CSB interaction works
repulsively ($+0.20$ MeV) and attractively ($-0.20$ MeV), 
respectively, in  $^7_{\Lambda}$He and $^7_{\Lambda}$Be.
As a result, our calculated values do not
reproduce  the  observed $B_{\Lambda}$s of 
$^7_{\Lambda}$He and $^7_{\Lambda}$Be.
Furthermore, in Ref. \citen{Hiyama09}, we pointed out that
the same phenomena was seen in 
the energy difference of the $T=1/2$
isodoublet $A=8$ hypernuclei ($^8_{\Lambda}$Li,
$^8_{\Lambda}$Be):
The agreement to the observed data of the energy difference
becomes worse by introducing the CSB $\Lambda -t(^3$He) interaction.

Let us discuss the energy difference of 
$^{10}_{\Lambda}$Be and $^{10}_{\Lambda}$B
using even-state CSB interaction employed in
$A=7$ hypernuclei.

First, in Fig. 6(a), we show the energy spectra 
of the  $A=10$ hypernuclei calculated without a CSB interaction.
The calculated $B_{\Lambda}$ values of $^{10}_{\Lambda}$Be and
$^{10}_{\Lambda}$B are 8.94 and 8.76 MeV, respectively.

Second, we turn on 
the even-state CSB interaction.
In Fig. 6(b), 
it is found that the CSB interaction works
repulsively by $+0.1$ MeV and attractively 
by $-0.1$ MeV in $^{10}_{\Lambda}$Be and $^{10}_{\Lambda}$B,
respectively.
This behavior is similar to the case of
$A=4$ and 7 hypernuclei.

Let us consider the energies of these $A=10$ hypernuclei more in detail.
In the case of $^{10}_{\Lambda}$Be,
the CSB interaction between the
$\Lambda$ and a valence neutron works repulsively and
the ground-state binding energy leads to
$B_{\Lambda}=8.83$ MeV, which is
 less bound by           
$0.1$ MeV than the value without the 
CSB effect.         
In $^{10}_{\Lambda}$B,                
the CSB interaction contributes attractively by 0.1 MeV,
 and  
the binding energy of the ground state                   
is $B_{\Lambda}=8.85$ MeV, which is close to the experimental data.
In order to see the CSB effect in the $A=10$
hypernuclei more clearly, 
let us evaluate
the difference between the  calculated $B_\Lambda$ values for
$^{10}_{\Lambda}$Be and $^{10}_{\Lambda}$B;
$\Delta B_{\Lambda}^{\rm cal}=
B_{\Lambda}^{\rm cal}(^{10}_{\Lambda}{\rm B})
-B_{\Lambda}^{\rm cal}(^{10}_{\Lambda}{\rm Be})
=-0.18$ MeV  without CSB,
which is in good agreement with the experimental value,
$\Delta B_{\Lambda}^{\rm exp}=
B_{\Lambda}^{\rm exp}(^{10}_{\Lambda}{\rm B})
-B_{\Lambda}^{\rm exp}(^{10}_{\Lambda}{\rm Be}) =-0.22 \pm 0.25 $ MeV. 
Switching on the even-state CSB interaction, the value obtained for 
$\Delta B_{\Lambda}^{\rm cal}=0.02$MeV moves away from
the central value of the data, $-0.22$ MeV.

%
In this way, we find that if we introduce 
a phenomenological $\Lambda N$ CSB interaction,
the binding energies of $A=7,8,10$ $\Lambda$ hypernuclei 
become inconsistent with the observed data.

We can also discuss the  CSB effects in
 $s$-shell and $p$-shell $\Lambda$ hypernuclei
from the experimental data.
The observed biding energy of 
$^4_{\Lambda}$He is larger by 0.35 MeV
than that of $^4_{\Lambda}$H.
Namely, it seems that $p \Lambda$ interaction 
is more attractive than $n \Lambda$ interaction
by CSB effect.
While, the observed binding energies of
$^7_{\Lambda}$He, $^8_{\Lambda}$Li and
$^{10}_{\Lambda}$Be are larger than
those of $^7_{\Lambda}$Be,
$^8_{\Lambda}$Be and $^{10}_{\Lambda}$B.
This means that the $n \Lambda$ interaction is
more attractive than the $p \Lambda$ interaction.

One possibility to solve this contradiction
is to re-investigate the experimental data, especially
those of $s$-shell $\Lambda$
hypernuclei, $^4_{\Lambda}$H, $^4_{\Lambda}$He.
In fact, it is planned to
measure the M1 transition from $1^+$ state to
the $0^+$ state in $^4_{\Lambda}$He
at E13 J-PARC project \cite{TamuraE13}
and to measure
$\Lambda$ separation energy of the
$0^+$ state in $^4_{\Lambda}$H at Maintz and JLab.

\begin{figure}[htb]
\begin{center}
\centerline{\includegraphics[width=12 cm,height=13 cm]
{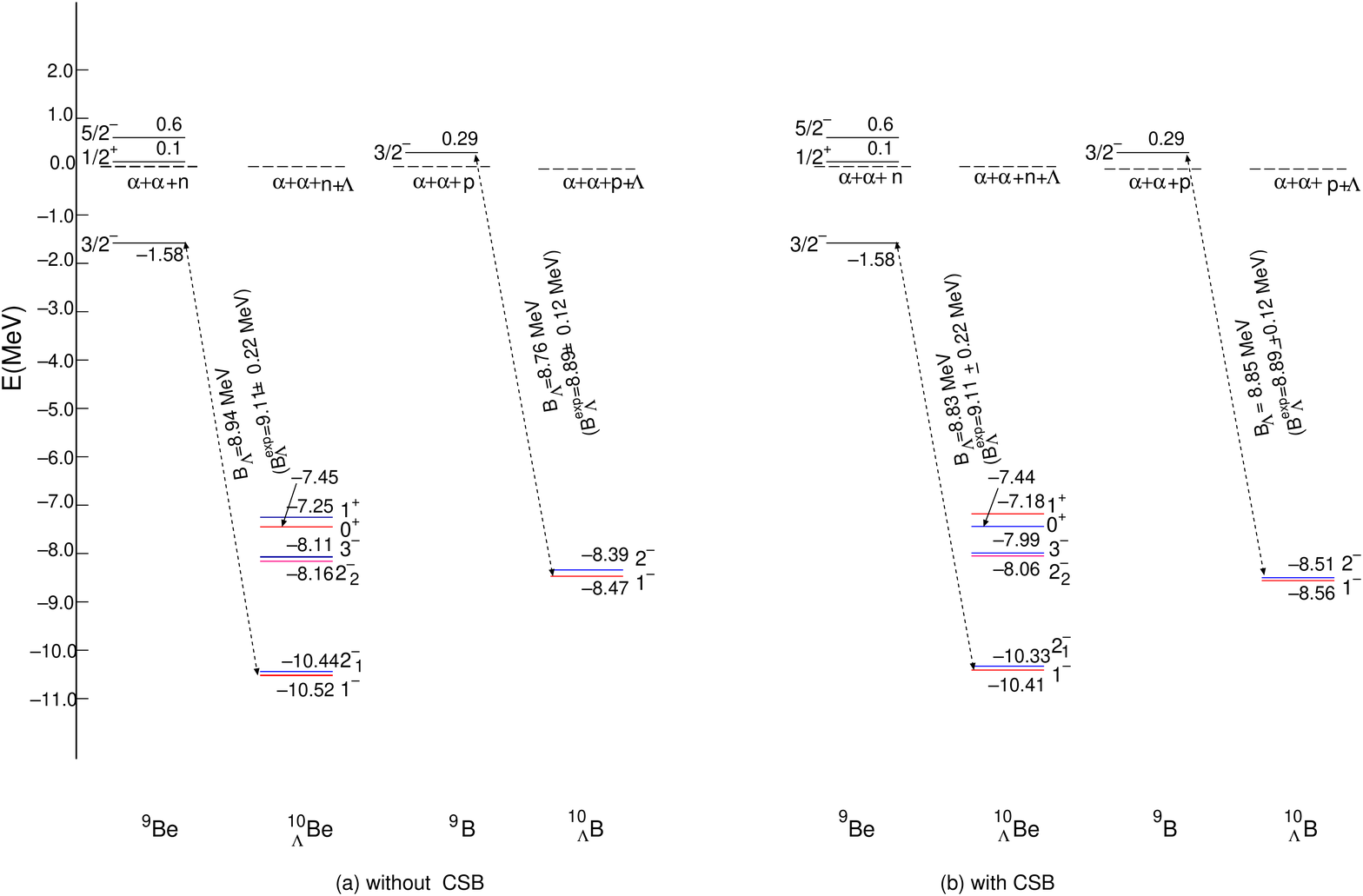}}
\label{fig:b10lcsb}
\caption{(color online).
Calculated energy levels of $^9$Be,
 $^{10}_{\Lambda}$Be, $^9$B, and $^{10}_{\Lambda}$B with
 spin-spin and spin-orbit $\Lambda N$ interactions.
 The even-state CSB potential is not included in the 
 calculated energies of $^{10}_{\Lambda}$Be and
 $^{10}_{\Lambda}$B of  (a),
 and included in  those of (b).
 The energies are measured from the particle breakup threshold.}
\end{center}
\end{figure}

\begin{figure}[htb]
\begin{center}
\centerline{\includegraphics[width=13 cm,height=9 cm]
{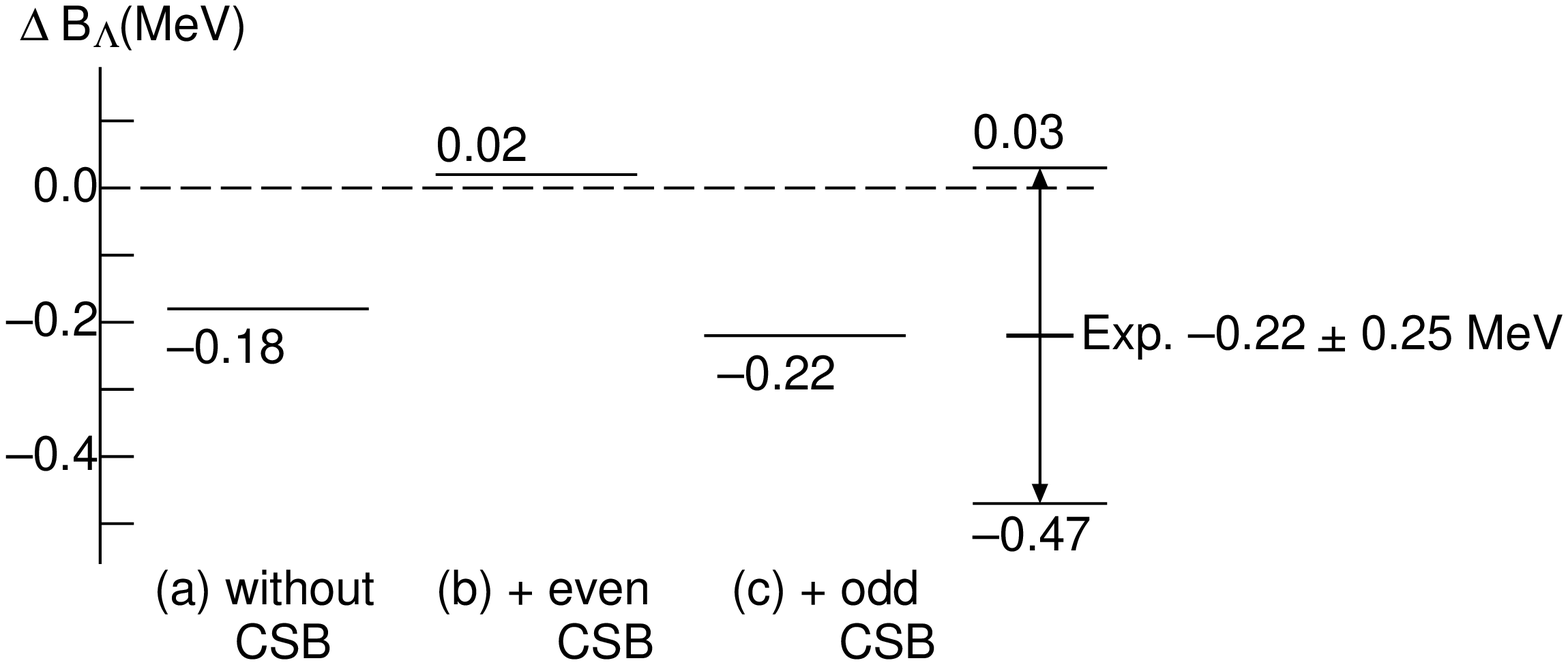}}
\label{fig:be10lcsb}
\caption{
Calculated energy difference of $^{10}_{\Lambda}$Be
and $^{10}_{\Lambda}$B, $\Delta B_{\Lambda}(
B_{\Lambda}(^{10}_{\Lambda}{\rm B})
-B_{\Lambda}(^{10}_{\Lambda}{\rm Be})$, (a) without CSB,
(b) with even-state CSB, and
(c) with both even- and odd-state
CSB interactions}
\end{center}
\end{figure}

One of candidates to solve the contradiction is simply
to introduce the odd-state CSB interaction with
opposite sign to the even-state CSB interaction.
The odd-state CSB interaction is negligible in
$s$-shell $\Lambda$ hypernuclei
but significant in $p$-shell $\Lambda$ hypernuclei.

In Ref.\citen{Hiyama09},
we pointed out that 
in order to reproduce the
data of these $A=8$ hyernuclei,
it was necessary the odd-state CSB interaction
with opposite sign of  that of the
even state CSB interaction \cite{Hiyama09}.

It is expected that such an odd-state CSB interaction plays the same role in the
$A=7$ and 10 hypernuclei.
Here, we show the results of $^{10}_{\Lambda}$Be
and $^{10}_{\Lambda}$B
without the CSB and with even-state CSB chosen to reproduce
the observed binding energies of
$^4_{\Lambda}$H and $^4_{\Lambda}$He, and  with even- and odd-state
CSB interactions. 
The odd-state CSB interaction is introduced with
opposite sign to that the even state part, 
but whose contributions in the
$^4_{\Lambda}$H and $^4_{\Lambda}$He are negligible.
Their potential parameters, strengths and ranges, are
fixed so as to reproduce our calculated
$B_{\Lambda}$ for $^7_{\Lambda}$He, 5.36 MeV.
The detailed potential parameters are mentioned
in the subsection 3.2.
%

\begin{table}
\begin{center}
\caption{
Calculated $\Lambda$ separation energies of
$A=7$ and 10 $\Lambda$ hypernuclei with and without
CSB interactions. The 
$B_{\Lambda}^{\rm cal}(^{10}_{\Lambda}{\rm Be})
-B_{\Lambda}^{\rm cal}(^{10}_{\Lambda}{\rm B})$
and 
$\Delta B_{\Lambda}^{\rm exp}=
B_{\Lambda}^{\rm exp}(^{10}_{\Lambda}{\rm Be})
-B_{\Lambda}^{\rm exp}(^{10}_{\Lambda}{\rm B})$
are listed here. 
}
\label{tab:CSB}
\begin{tabular}{ccccc}
\hline \hline   \\
  &   & with & with & \\
  &without CSB   &even-state CSB  &even+odd-state CSB
  &Exp. \\
  \hline \\
$B_{\Lambda}(^7_{\Lambda}$He) &5.36 &5.16 &5.36 
&$5.68 \pm 0.03 \pm 0.25$ \cite{Hashimoto2011,Nakamura2012}
\\
$B_{\Lambda}(^7_{\Lambda}$Li)  &5.28 &5.29 &5.28 &5.26 \\
$B_{\Lambda}(^7_{\Lambda}$Be)  &5.21 &5.44 &5.27 &5.16$\pm 0.08$ \\
\hline \\
$B_{\Lambda}(^{10}_{\Lambda}$Be) &8.94 &8.83 &8.96 &$9.11 \pm 0.22$
\\
$B_{\Lambda}(^{10}_{\Lambda}$B)  &8.76 &8.85 &8.74 &$8.89 \pm 0.12$ \\
$\Delta B_{\Lambda}^{\rm cal}$ &0.18 &$-0.02$ &0.22 &$0.22 \pm 0.25$ \\
\hline \hline
\end{tabular}
\end{center}
\end{table}

Next, we apply the strong  odd-state CSB interaction
to level structures of
$^{10}_{\Lambda}$Be and $^{10}_{\Lambda}$B.
The calculated $\Lambda$-separation energies of
the $1^-$ states for $^{10}_{\Lambda}$Be and $^{10}_{\Lambda}$B
are $8.96$ MeV and $8.74$ MeV, respectively.
Then, 
$\Delta B_{\Lambda}^{\rm cal}= -0.22$ MeV,
which is in good agreement  with
$\Delta B_{\Lambda}^{\rm exp} =-0.22 \pm 0.25 $.
The biding energies of $A=7$ and 10 $\Lambda$ hypernulclei
with and without CSB interaction are listed in
Table 2.

Three results of $A=10$ hypernuclei
are summarized in Fig. 7.
Three results shown by solid lines are found to be within 
the experimental  error bars.
We see  deviation by 200 keV
in the $\Delta B_\Lambda^{\rm exp}$ with and without the CSB 
interaction.
Then, if  high resolution experiments can provide us 
new data for $^{10}_{\Lambda}$Be
and $^{10}_{\Lambda}$B within 100 keV accuracy
in the future ,
we can obtain information about
the CSB interaction.

As shown in Table 2 and Fig.7,
the biding energies of $A=7$ and 10 $\Lambda$
hypernuclei without CSB interaction
reproduce the all data.
However, the even-state CSB interaction
which reproduce the data of $s$-shell $\Lambda$
hypernuclei, $^4_{\Lambda}$H and $^4_{\Lambda}$He
leads to inconsistency of the
biding energies of $p$-shell $\Lambda$
hyernuclei.
As a trial, then, if we introduce a strong odd-state CSB interaction
with opposite sign of even-state CSB interaction,
we could reproduce the observed biding energies of
$A=7$ and 10 $\Lambda$ hypernuclei.
However, there still remains a room to discuss the validity 
of such a strong odd-state CSB interaction.
For the CSB effect in light $\Lambda$ hypernuclei,
it is necessary to re-investigate experimental data of
$s$-shell $\Lambda$
hypernuclei and $p$-shell $\Lambda$ hypernuclei
as mentioned before.  
In fact, it is planned to
measure the M1 transition from $1^+$ state to
the $0^+$ state in $^4_{\Lambda}$He
at E13 J-PARC project \cite{TamuraE13}
and to measure
$\Lambda$ separation energy of the
$0^+$ state in $^4_{\Lambda}$H at Maintz and JLab.
From these measurements, we could conclude whether or not
there exist in CSB effect in the
binding energies of $A=4$ hypernuclei. 
We need wait for these data.

\section{Summary}

We study the structure of hypernuclear isodoublet
$^{10}_{\Lambda}$B and $^{10}_{\Lambda}$Be
within the framework of $\alpha +\alpha +\Lambda +N$
four-body model.
In this model, it is important that all 
two-body interactions among subunits (two $\alpha$'s,
$\Lambda$ and $N$) are chosen so as to reproduce
the binding energies of all subsystems composed of
two and three subunits.
The  $\Lambda N$ interaction,
which simulates $\Lambda N$ scattering phase shifts
of NSC97f, are adjusted so as to
reproduce the observed data for the
spin-doublets states, $0^+$-$1^+$ and
$1/2^+$-$3/2^+$, of $^4_{\Lambda}$H and
$^7_{\Lambda}$Li, respectively.
Before discussing major conclusion,
we comment on our general viewpoint for effective interactions 
used in our cluster-model analyses.
Our basic assumption in this work
is that the $\Lambda N-\Sigma N$ coupling interaction
can be renormalized into the
$\Lambda N-\Lambda N$ interaction effectively.
It should be noted that our renormalizations into effective $\Lambda N$ interactions are made so as to reproduce experimental values of binding energies of subunits such as $\Lambda N$, 
$\Lambda \alpha$, $\Lambda \alpha \alpha$ and so on.
Here we emphasized that
the validity of nuclear models and effective interactions in them should be based on the consistency with experimental data:
In our cluster-model approach, the experimental data of above hypernuclei are reproduced systematically with use of our
effective  interactions.

The main conclusions can be summarized as follows:

(1) We calculated spin-doublet states of  $1^-$-$2^-_1$
in $^{10}_{\Lambda}$B whose measurement  was obtained in
 BNL930 \cite{BNL930}.
The calculated splitting energy is 0.08 MeV. 
This small value is less than the  0.1 MeV precision 
for detecting the M1 transition from the $2^-$ to the $1^-$ state,
which is consistent with the experimental fact of
no observed $\gamma$-ray. 
Furthermore, we calculated the spin-doublets, $1^-$-$2^-_1$
, $2^-_2$-$3^-$ and $0^+$-$1^+$ state, of $^{10}_{\Lambda}$Be.
The measurement for $^{10}_{\Lambda}$Be was done at
JLab and the analysis is in progress.
The energy spillitings of these states are predicted to be
0.08 MeV, 0.05 MeV and 0.2 MeV, respectively.
Then, it would be difficult to observe the energy 
splittings for the negative parity
which are produced by $(K^-,\pi^-)$
experiment, 
although these energy
splittings would be helpful for
extracting information about
the $\Lambda N$ spin-dependent components.

(2) The effect of the glue-like role of the 
$\Lambda$ particle can be demonstrated in 
$^{10}_{\Lambda}$B and $^{10}_{\Lambda}$Be.
The ground state of $^9$B is a resonant state.
Due to the presence of the $\Lambda$ particle,
the ground state of 
the resultant hypernucleus $^{10}_{\Lambda}$B
becomes bound by about 2.0 MeV below the
$^9_{\Lambda}$Be$+p$ threshold.
When the $\Lambda$ particle is added to
the bound ground state of $^9$Be,
the corresponding state of $^{10}_{\Lambda}$Be
becomes bound more deeply by about 4 MeV below the
$^9_{\Lambda}$Be$+n$ threshold.
Furthermore, by adding the $\Lambda$ particle to
the resonant state of $^9$Be, $1/2^+$ and $5/2^-$,
the corresponding states of $^{10}_{\Lambda}$Be
become bound.
Especially, we find that the order
of the $3^-$, $2^-_2$, $0^+$ and $1^+$
states is reversed from $^9$Be to
$^{10}_{\Lambda}$Be.
From the calculated values of the rms radii $\bar{r}_{\alpha -\alpha}$ 
and $\bar{r}_{\alpha -\Lambda}$ of $^9$Be and $^{10}_{\Lambda}$Be,
we find the shrinkage effect due to
the addition of $\Lambda$ to the core nucleus.

Such an effect was already confirmed by 
at KEK-E419 experiment~\cite{Tanida01}.
Another interesting feature seen in our result is
the three-layer structure of
the matter distributions in
isodoublet hypernuclear states,
being composed of
a 2$\alpha $ core, a $\Lambda$, and
a nucleon.
Also, we have neutron halo structures  for the 
$0^+$ and $1^+$ states.

(3) The charge symmetry breaking effect in
$^{10}_{\Lambda}$Be and $^{10}_{\Lambda}$B
are investigated quantitatively on
the basis of the phenomenological
CSB interaction, which describe the
experimental energy difference 
between $B_{\Lambda}(^4_{\Lambda}$H)
and $B_{\Lambda}(^4_{\Lambda}$He),
$\Delta_{\rm CSB}$.
We introduce $\Delta B_{\Lambda} = B_{\Lambda}(^{10}_{\Lambda}{\rm Be})
-B_{\Lambda}(^{10}_{\Lambda}{\rm B})$.
And we  obtained $\Delta B_{\Lambda}^{\rm cal}$
to $-0.02 \sim 0.22$ MeV
without and with the CSB interaction,
which agree with the observed $\Delta B_{\Lambda}^{\rm exp}$
within the large error bar.

In order to elucidate CSB effects in light
hypernuclei, it is necessary to
have precise data for
$^4_{\Lambda}$H, $^4_{\Lambda}$He,
$^7_{\Lambda}$He, $^7_{\Lambda}$Li ($T=1$), $^7_{\Lambda}$Be,
$^{10}_{\Lambda}$Be, and
$^{10}_{\Lambda}$B.
The calculated $\Lambda$ separation energies of
$p$-shell hypernuclei became inconsistent with
the observed data when we use the even-state
CSB interaction to reproduce the observed data of
$s$-shell $\Lambda$ hypernuclei of
$^4_{\Lambda}$H and $^4_{\Lambda}$He.

In this contradictory situation, one possibility
is to re-investigate the experimental data, especially
those of $^4_{\Lambda}$H, $^4_{\Lambda}$He.
At J-PARC, it is planned to measure the M1 transition from
the $1^+$ state to the $0^+$ state in $^4_{\Lambda}$He
at E13 J-PARC project \cite{TamuraE13}
and to measure $\Lambda$ separation energy of the
$0^+$ state in $^4_{\Lambda}$H at Mainz and JLab.
From these measurements, we can investigate
the interesting issue of whether or not
there is a CSB effect in the
binding energies of $^4_{\Lambda}$He and
$^4_{\Lambda}$H.
These experimental results have to affect
the CSB effect in $p$-shell $\Lambda$ hypernuclei.

As a working assumption to explain the CSB effects
both in A=4 and $p$-shell systems, we have introduced
the extremely-repulsive odd-state CSB interaction
to cancel out the even-state CSB contributions.
Even if this assumption works well, it is an open
problem to elucidate physical reality for it.
In order to find the effects of the odd-state CSB
in $A=10$ hypernuclei,
we need data with 0.1 MeV resolution.
In the case of  $^{10}_{\Lambda}$B,
we propose to  perform  the  experiment
$^{10}$B $(K^-, \pi^-)^{10}_{\Lambda}$B
at J-PARC in the future.
In the case of $^{10}_{\Lambda}$Be,
the experiment of $^{10}$B $(e,e'K^+) ^{10}_{\Lambda}$Be
at JLab was done and analysis is in progress.
We hope to have the $\Lambda$ separation energy
for this hypernucleus with 0.1 MeV resolution.


%
%
\section*{Acknowledgments}
The authors thank Professors O. Hashimoto,  
H. Tamura, and S. N. Nakamura,  T. Motoba,
B.\ F.\ Gibson, and 
D. J. Millener for helpful discussions.
This work was supported by a Grants-in-Aid for
Scientific Research from Monbukagakusho of Japan(21540288, 20105003).
The numerical calculations were performed 
at the Yukawa Institute Computer Facility HITACHI-SR16000
and KEK-SR16000.

\end{document}